\documentclass[twocolumn,showpacs,epsfig,superscriptaddress]{revtex4}

\usepackage{graphicx}
\usepackage{amssymb}
\usepackage{amsmath}
\usepackage{color}
\usepackage{bm}
\usepackage{physics}
\usepackage{changes}

\newcommand{\figsizethree}{0.8}

\makeatletter
\let\Changes@Markup@Deleted\@gobble
\makeatother

\begin{document}

\title{Emergent localized states at the interface of a twofold $\mathcal{PT}$-symmetric lattice}
\author{Jung-Wan Ryu}
\affiliation{Center for Theoretical Physics of Complex Systems, Institute for Basic Science (IBS), Daejeon 34126, South Korea}
\author{Nojoon Myoung}
\affiliation{Department of Physics Education, Chosun University, Gwangju 61452, Republic of Korea}
\author{Sungjong Woo}
\affiliation{Center for Theoretical Physics of Complex Systems, Institute for Basic Science (IBS), Daejeon 34126, South Korea}
\author{Ara Go}
\affiliation{Center for Theoretical Physics of Complex Systems, Institute for Basic Science (IBS), Daejeon 34126, South Korea}
\author{Sang-Jun Choi}
\email{aietheia@ibs.re.kr}
\affiliation{Center for Theoretical Physics of Complex Systems, Institute for Basic Science (IBS), Daejeon 34126, South Korea}
\author{Hee Chul Park}
\email{hcpark@ibs.re.kr}
\affiliation{Center for Theoretical Physics of Complex Systems, Institute for Basic Science (IBS), Daejeon 34126, South Korea}
\date{\today}

\begin{abstract}
We consider the role of non-triviality resulting from a non-Hermitian Hamiltonian that conserves twofold $\mathcal{PT}$-symmetry assembled by interconnections between a $\mathcal{PT}$-symmetric lattice and its time reversal partner.
Twofold $\mathcal{PT}$-symmetry in the lattice produces additional surface exceptional points that play the role of new critical points, along with the bulk exceptional point. We show that there are two distinct regimes possessing symmetry-protected localized states, of which localization lengths are robust against external gain and loss. The states are demonstrated by numerical calculation of a quasi-1D ladder lattice and a 2D bilayered square lattice.
\end{abstract}
\maketitle
\narrowtext

Parity--time ($\mathcal{PT}$) symmetric systems exhibit a phase transition through spontaneous symmetry breaking, from an unbroken $\mathcal{PT}$-symmetric phase that keeps the eigenenergy real in non-Hermitian systems, to a broken phase that contains complex conjugate energies \cite{Ben98}. As the general characteristic of non-Hermitian systems, non-Hermiticity develops from imaginary phase accumulation and imaginary potential due to an imbalance of particle/energy flow where $H^{\dagger}\neq H$~\cite{Hat96, Isr19, Elg07, Rue10}. $\mathcal{PT}$-symmetry is protected in non-Hermitian systems with a balance of energy gain and loss represented by the commutation relation $[H, \mathcal{PT}] = 0$, where $H$ is a Hamiltonian and $\mathcal{PT}$ is a combination of parity and time-reversal symmetry operators. A wide range of $\mathcal{PT}$-symmetric systems have been explored over several fields, including optics \cite{Elg07, Mus08, Mak08, Kla08, Guo09, Rue10, Reg12}, electronic circuits \cite{Sch11}, atomic physics \cite{Jog10}, and magnetic metamaterials \cite{Laz13}.
Phase transitions in these systems occur via exceptional points (EPs), which are degenerate points of eigenenergies in non-Hermitian systems that generate a M{\"o}bius strip structure of eigenenergies in parametric space because of the square-root branching property of the singular point \cite{Kat96, Hei12}. Such a topological structure has been reported in microwave experiments \cite{Dem01, Dem03}, optical microcavities \cite{Lee09}, and a chaotic exciton-polariton billiard \cite{Gao15}.

In non-Hermitian systems, states localized at the edges, interfaces, and defects have recently attracted considerable attention not only in fundamental studies such as topology and symmetry but also in applications to quantum technologies.
Non-Hermitian flat bands generate localized zero modes analogous to Majorana zero modes in condensed matter physics, of which real energies are zero but imaginary energies are non-zero \cite{Li15, Li18}. These non-Hermitian zero modes (NHZMs) are protected by non-Hermitian particle--hole (NHPH) symmetry, which is also called charge-conjugate symmetry, where $\epsilon_i = - \epsilon_j^*$ with $\mathrm{Re}(\epsilon_i) = 0$ if $i=j$ \cite{LGe17, Qi18}. 
It has been shown that NHPH symmetry generates topological defect modes at the interface between non-Hermitian lattices based on topologcially trivial Hermitian lattices \cite{Mal15, Mal18}. Otherwise, topologically protected NHZMs have been proposed in non-Hermitian lattices based on topologically non-trivial Hermitian lattices, such as the Su--Schrieffer--Heeger model \cite{Sch13, Pol15, Zeu15, Zha18, Par18, Su79}.
In addition to NHZMs, non-Hermitian bound states (NHBSs) are protected by $\mathcal{PT}$-symmetry, where $\epsilon_i = \epsilon_j^*$ with $\mathrm{Im}(\epsilon_i) = 0$ if $i=j$ \cite{Wei17, Ni18}.
Anomalous localized edge states in non-Hermitian lattice models have been attributed to a winding number around an EP in momentum space \cite{Lee16, Ley17, Mar18}; localized states also exist at the interface between two lattices with the same topological order but with distinct quantum phases, such as unbroken and broken $\mathcal{PT}$-symmetric phases \cite{Zha15, Pan18}.

A $\mathcal{PT}$-symmetric system can be realized by combining the even-parity of the real potential and the odd-parity of the imaginary potential with respect to the $\mathcal{PT}$-symmetric axis. Pairing the systems to be time-reversal partners induces twofold $\mathcal{PT}$-symmetry, which generates two different symmetric axes and guarantees two EPs. 
Such multiple EPs by multifold $\mathcal{PT}$-symmetry as well as single EPs by simple $\mathcal{PT}$-symmetry have been studied both theoretically and experimentally \cite{LGe14, Liu16, Din16}. 
In this work, we introduce a lattice containing twofold $\mathcal{PT}$-symmetry to study how interplay between non-Hermiticity and bulk symmetry affects the wavefunctions of the lattice. 
We find two pairs of interface states that decay exponentially in the space distinguished by the bulk states.
Particularly, we focus on robust localized states (NHZMs and NHBSs) of which localization length is unaffected by variation of external gain/loss in two distinct regimes between the bulk EP and surface EPs.

\begin{figure}
\begin{center}
\includegraphics[width=0.9\linewidth]{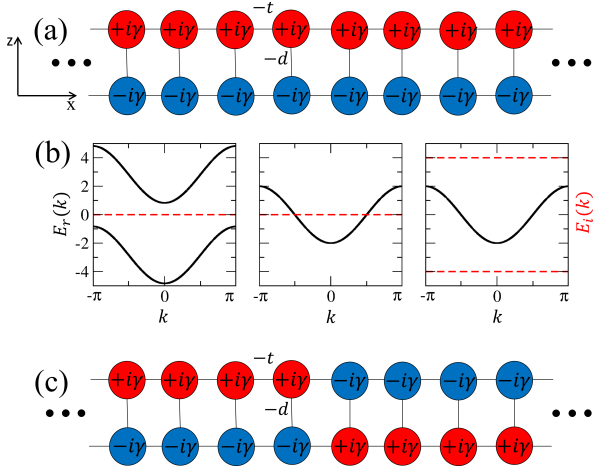}
\caption{(color online).
(a) Sketch of a $\mathcal{PT}$-symmetric ladder lattice with gain (red) and loss (blue) on the upper and lower lattices, respectively. (b) Real (black solid line) and imaginary (red dashed line) parts of the complex energy bands of a $\mathcal{PT}$-symmetric ladder lattice for $\gamma<\gamma_b$ (left), $\gamma=\gamma_b$ (middle), and $\gamma>\gamma_b$ (right). The parameters are $d=3$ and $t=1$. (c) Sketch of the intertwined $\mathcal{PT}$-symmetric ladder lattice considered here.
}
\label{fig1}
\end{center}
\end{figure}

Let us prepare the $\mathcal{PT}$-symmetric ladder lattice with a basis of two sites (see Fig.~\ref{fig1} (a)). Non-Hermiticity is adopted by respective gain and loss at the two sites in the unit cell, a balanced amount of which preserves $\mathcal{PT}$-symmetry. A series of such unit cells forms a ladder lattice with inter-cell hopping.
The lattice Hamiltonian for this system reads $\hat{H}_0=\sum_{l=-\infty}^{\infty}\hat{h}_l$, where
\begin{eqnarray}
\hat{h}_l=c_l^{\dagger} h c_l + c_{l}^{\dagger} T^{\dagger} c_{l+1} + c_{l+1}^{\dagger} T c_l
\label{eq:spt}
\end{eqnarray}
and $c_{l}^{\dagger}\equiv(\ket{l,A},\ket{l,B})$. The matrix $h\equiv i\gamma\sigma_z-d\sigma_x$ describes the non-Hermicity with $\gamma$ and the intra-cell hopping with $-d$, while $T\equiv-t\sigma_0$ describes the inter-cell hopping where $\sigma_{x,z}$ are the Pauli matrices and $\sigma_0$ is the $2\times 2$ identity matrix.

\begin{figure}
\begin{center}
\includegraphics[width=0.9\linewidth]{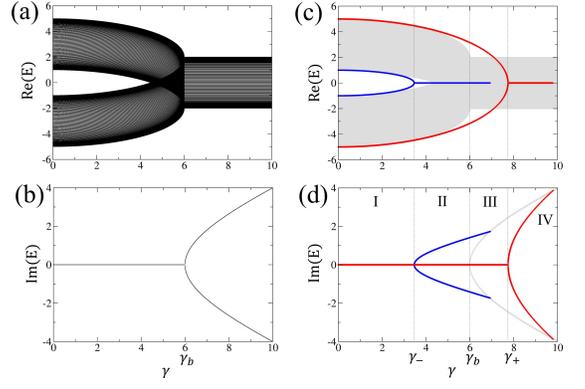}
\caption{(color online).
(a) Real and (b) imaginary parts of the complex eigenenergies of Fig. \ref{fig1} (a) as a function of $\gamma$. $\gamma_b$ is the exceptional point related to the $\mathcal{PT}$-symmetric phase transition. 
(c) Real and (d) imaginary parts of the complex eigenenergies of Fig. \ref{fig1} (c) as a function of $\gamma$. The gray shaded region represents the bulk state eigenenergies, and the red and blue lines indicate the eigenenergies of the interface states. $\gamma_b$ is the bulk exceptional point and $\gamma_{\pm}$ are the surface exceptional points for the interface states.
The parameters are $d=3$ and $t=1$, and the number of unit cells is $N=400$.
}
\label{fig2}
\end{center}
\end{figure}

The matrix representation of the Hamiltonian for a single ladder lattice in momentum space is written as $H_0(q)=h-2t\cos{q}\sigma_0$ through the Bloch wavefunction $\ket{\psi(q)}\equiv\sum_n e^{inq}c_n^{\dagger}$, where $q$ is the momentum vector. The (non-normalized) eigenstates are $\psi_{\eta}(q) = (-i \gamma/2 - \eta \sqrt{d^2 - \left(\gamma/2\right)^2}, d)^T$, which are independent of $q$, where $\eta=\pm$ assigns the upper/lower bands. The corresponding eigenvalues of the Hamiltonian are $E_\eta(q) =-2t \cos{q} +\eta \sqrt{d^2 - \left(\gamma/2\right)^2}$, which are complex energy bands as $E(k)=E_r(k) + i E_i(k)$ according to different $\gamma$ in Fig.~\ref{fig1} (b). One may notice that there exists a phase transition between unbroken and broken $\mathcal{PT}$-symmetric phases through the EP. At this EP, two complex energy bands merge into a single energy band, as shown in middle panel of Fig.~\ref{fig1} (b); otherwise, the real and imaginary bands are separate.
The real and imaginary parts of the complex spectrum for a finite-sized $\mathcal{PT}$-symmetric ladder lattice with $400$ unit cells are numerically calculated as a function of $\gamma$ in Fig.~\ref{fig2} (a) and (b), respectively.
Results show a collective $\mathcal{PT}$-symmetric phase transition at the bulk EP $\gamma_b=2d$. For $\gamma<\gamma_b$, the energy spectrum pair is real in spite of non-Hermiticity, with each part attracting each other up to the bulk EP unlike the Hermitian cases, as shown in Fig.~\ref{fig2} (a). For $\gamma>\gamma_b$, meanwhile, the energy spectrum pairs are complex conjugates, and repel each other on the imaginary energy plane with increasing $\gamma$ from the bulk EP, as shown in Fig.~\ref{fig2} (b). 
We note that the two bands are separable at any value of $\gamma$ except the bulk EP, because there is no band crossing for any $q$ \cite{She18}.

The twofold $\mathcal{PT}$-symmetric ladder lattice can be formed by merging the edges of the prepared lattice and its time-reversal partner to create an interface in a process called intertwinement (see Fig.~\ref{fig1} (c)).
The Hamiltonian operator of this intertwined lattice consists of
\begin{eqnarray}
\hat{H}=\hat{H}_L+\hat{H}_R+\hat{H}_c,
\end{eqnarray}
where $\hat{H}_L=\sum_{l=-\infty}^{-1}\hat{h}_{l}$ is a Hamiltonian for the left semi-infinite non-Hermitian lattice, and $\hat{H}_R=\mathcal{T}\sum_{l=0}^{\infty}\hat{h}_{l}\mathcal{T}^{-1}$ is a Hamiltonian for the right semi-infinite lattice with complex conjugate operator $\mathcal{T}$ as a time-reversal operator. A coupling Hamiltonian $\hat{H}_c=c_0^{\dagger} T c_{-1}+h.c.$ connects the two time-reversal partners. Intertwinement introduces additional $\mathcal{PT}$-symmetry with respect to the $z$-axis and additional EPs according to the symmetry.
The energy spectrum of the finite-sized twofold $\mathcal{PT}$-symmetric ladder lattice is treated by using a tridiagonal Hamiltonian. As shown in Fig.~\ref{fig2} (c) and (d), the four emergent states (red/blue lines) not only show the $\mathcal{PT}$-symmetric phase transition but also separation from the bulk states, while the bulk states are coincident with the states of the simple $\mathcal{PT}$-symmetric lattice.

\begin{figure}
\begin{center}
\includegraphics[width=0.9\linewidth]{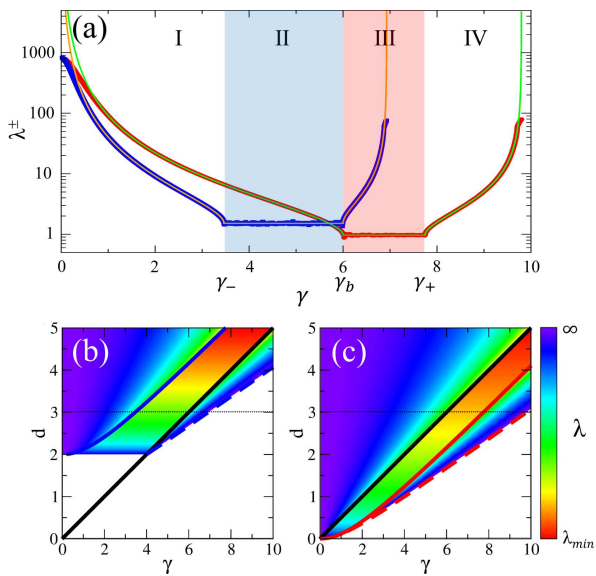}
\caption{(color online).
(a) Localization lengths of odd (blue line) and even (red line) interface states using the same parameters as Fig. \ref{fig2}. The NHZMs and NHBSs are robust against $\gamma$ in the blue and red shaded regimes II and III, respectively. The orange and green lines represent the analytic solutions of localization lengths of the odd and even interface states, respectively. (b, c) Phase diagrams for the localization lengths of odd and even states, respectively. Blue solid, black solid, and blue dashed lines in (b) indicate $\gamma_-$, $\gamma_b$, and $\gamma_{-b}$, respectively. Black solid, red solid, and red dashed lines in (c) indicate $\gamma_b$, $\gamma_+$, and $\gamma_{+b}$, respectively. No localized states exist in the white areas. The black dotted lines indicate the localization lengths shown in (a).
}
\label{fig3}
\end{center}
\end{figure}

The emergent states are exponentially localized at the interface, of which localization lengths defined by the exponential decay of the wavefunctions $\psi\sim e^{-x/\lambda}$ as a
function of $\gamma$ are shown in Fig. \ref{fig3} (a).
Two regimes having constant localization lengths are indicated by the shaded regions in Fig.\ref{fig3} (a).
In the blue-shaded regime $\gamma_- <\gamma<\gamma_b$, the localized interface states conserve NHPH symmetry as NHZMs, and in the red-shaded regime $\gamma_b<\gamma<\gamma_+$, the localized interface states conserve $\mathcal{PT}$-symmetry as NHBSs.
Otherwise, the localized states are regarded as defect states in a band gap.
The localization lengths of the NHZMs and NHBSs are independent of the non-Hermiticity control parameter $\gamma$, which is related to the interplay between the two different $\mathcal{PT}$-symmetric phases of interface and bulk states on the lattice protecting twofold $\mathcal{PT}$-symmetry. The NHZMs appear when the bulk and interface states have unbroken and broken phases, respectively, while conversely the NHBSs appear when the bulk and interface states have broken and unbroken phases, respectively. Otherwise, the interface and bulk states have the same unbroken or broken phases.
The phase diagrams in Fig.~\ref{fig3} (b) and (c) show the localization lengths of the interface states of NHZMs and NHBSs as a function of intra-cell hopping $d$ and non-Hermiticity parameter $\gamma$, respectively.
When $d < 2t$, we can recognize that there exist only a pair of interface states, including NHBSs, by comparing the phase diagrams in which the white areas indicate forbidden regimes of localized states.
We note that, while invisible in the energy spectra in Fig. \ref{fig2} (c) and (d), phase transitions appear for the localized states at the bulk EP, as shown in Fig. \ref{fig3} (a).

The interface states can be obtained through the Schr\"odinger equation $H\ket{\Phi}=E\ket{\Phi}$. Let us consider a linear combination of the simple ladder lattice eigenstates to solve the intertwined lattice [see Supplemental Materials]. The ansatz of the wavefunction in the system is $\ket{\Phi}=\ket{\Phi}_L+\ket{\Phi}_R=\sum_{\eta=\pm}\lbrace\alpha_{\eta}\ket{\psi(q_{\eta})}_L+\beta_{\eta}\ket{\psi(-q_{\eta})}_R\rbrace$. The matching condition of the counter-propagating wavefunctions at the interface is
\begin{eqnarray}
2d^2\sin{q_+}\sin{q_-}+\left(\gamma\over 2\right)^2\cos{(q_+ + q_-)}=\left(\gamma\over 2\right)^2,
\end{eqnarray}
where $\cos{q_{\eta}}=\frac{-E+\eta\sqrt{d^2-(\gamma/2)^2}}{2}$ is the momentum vector corresponding to each energy band, with all energy scales being dimensionless by $t$ from here. This condition provides exact solutions for the interface states as
\begin{eqnarray}
\label{epm}
E^{\pm}_{\eta}=\eta\sqrt{\left(1\pm\frac{2}{d}\right)\left(d^2\pm 2d-\left(\frac{\gamma}{2}\right)^2\right)},
\end{eqnarray}
which are eigenenergies of the two paired states separated from the bulk bands, where the index $\pm$ indicates the even/odd interface states as the red/blue curves in Fig.~\ref{fig2} (c) and (d), respectively. It is easy to see that the two surface EPs are $\gamma_{\pm}=2\sqrt{d^2\pm 2d}$ and that the interface states undergo phase transitions from unbroken to broken $\mathcal{PT}$-symmetric phases.
There is no interface state in the Hermitian limit, i.e., $\gamma = 0$, and the interface states with real eigenenergies separate from the bulk states as $\gamma$ increases.
The even and odd interface states finally merge into the bulk states at $\gamma_{+b} = 2 \sqrt{2} \sqrt{d^2 + d}$ and $\gamma_{-b} = 2 \sqrt{2} \sqrt{d^2 - d}$, respectively.

To analyze the characteristics of the emergent interface states, we introduce complex momentum vectors of the states as $q^{\pm}_{\eta}=k^{\pm}_{\eta}+i\kappa^{\pm}_{\eta}$, where $k^{\pm}_{\eta}$ and $\kappa^{\pm}_{\eta}$ are real and ${\pm}$ indicates the even/odd state. 
The dispersion relation is generalized by complex momentum $\cos(k+i\kappa)=\cos{k}\cosh{\kappa}-i\sin{k}\sinh{\kappa}$.
By substituting the energies of the interface states as found in Eq.~(\ref{epm}) for the energy in the dispersion relation,
\begin{eqnarray}
\cos{q_{\eta}^{\pm}}=\frac{-E_{\eta}^{\pm}+\eta\sqrt{d^2-(\gamma/2)^2}}{2},
\end{eqnarray}
we can formulate localization lengths as $\lambda^{\pm} = 1 / \kappa^{\pm}_{\eta}$ at the interfaces of the intertwined lattice with respect to the distinct regimes of control parameter $\gamma$ [see Supplemental Materials]. 
For the NHZMs, we can evaluate the inverse localization length and wave number for oscillation as follows,
\begin{eqnarray}
\cosh{\kappa_{\eta}^{-}}=\sqrt{\frac{d}{2}},~\cos{k_{\eta}^{-}=-\eta\sqrt{\frac{d^2-(\gamma/2)^2}{2d}}}.
\label{odd_ll}
\end{eqnarray}
In case of the NHBSs, the inverse localization length and wave number of the localized states are
\begin{eqnarray}
\cosh{\kappa_{\eta}^{+}}=\sqrt{\frac{d}{2}+1},~\cos{k_{\eta}^{+}=\eta\sqrt{\frac{d^2+2d-(\gamma/2)^2}{2d}}}.
\label{even_ll}
\end{eqnarray} 
The localization lengths $1/\kappa_{\eta}^{\pm}$ from Eq.~(\ref{odd_ll}) and Eq.~(\ref{even_ll}) agree well with numerical results, as shown in Fig.~\ref{fig3} (a).

\begin{figure}
\begin{center}
\includegraphics[width=0.9\linewidth]{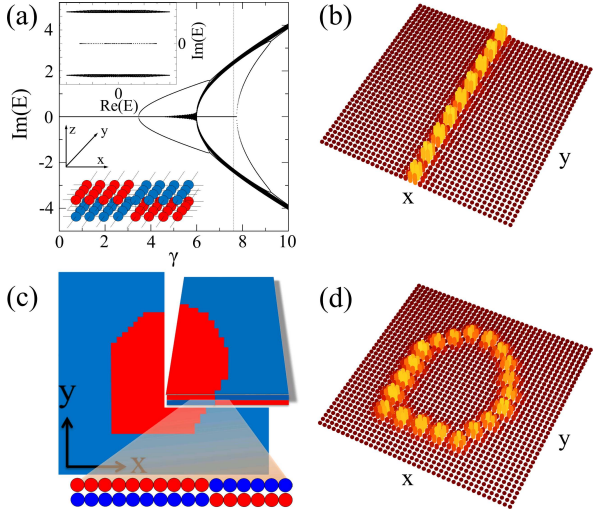}
\caption{(color online).
(a) Imaginary parts of complex eigenenergies as a function of $\gamma$ in the 2D ladder lattice. The upper inset plots the eigenenergies on the complex plane for $\gamma=7.6$, and the lower inset illustrates the 2D lattice with a 1D straight interface along the $y$-axis.
(b) Intensity distribution of a selected lossless NHBS for $\gamma=7.6$.
(c) Top view of the 2D lattice schematic with an arbitrarily shaped closed interface, with a cut-view showing the interface. The upper and lower layers have opposite profiles of gain and loss.
(d) Intensity distribution of a selected lossless NHBS for $\gamma=6.5$.
The parameters in this figure are $d=3$ and $t_{x}=t_{y}=1$, and the number of unit cells is $40 \times 40$.
}
\label{fig4}
\end{center}
\end{figure}

It is meaningful to understand the surface EPs through the eigenenergies of the twofold $\mathcal{PT}$-symmetric four sites containing self-energy, which have the same symmetries as our model. The effective Hamiltonian reads 
\begin{eqnarray}
H_{eff}=\tau_0\otimes(-d\sigma_x+i\frac{\gamma'}{2}\sigma_z)+\tau_x\otimes(-\sigma_x),
\end{eqnarray} 
where $\tau_i$ is the Pauli matrix with identity matrix $\tau_0$ for inter-cell hopping. The Hamiltonian is block diagonalized by means of unitary matrix $L=(\tau_x+\tau_z)\otimes\sigma_0/\sqrt{2}$. One of the blocks is $H_+=i\gamma'/2\sigma_z-(d+1)\sigma_x$ with eigenvalues $E_{\eta}^{+}=\eta\sqrt{(d+1)^2-(\gamma'/2)^2}$, and the other block is $H_-=i\gamma'/2\sigma_z-(d-1)\sigma_x$ with eigenvalues $E_{\eta}^{-}=\eta\sqrt{(d-1)^2-(\gamma'/2)^2}$. We can see two different critical values of $\gamma_{\pm}'$ with respect to the separated blocks. Suppose that the on-site energy is renormalized by $\gamma'^2=\gamma^2 + 4$ in terms of the self-energy of the semi-infinite leads; then the block diagonal matrix has two EPs, $\gamma_{\pm}=2\sqrt{d^2\pm 2d}$, which are the same as the surface EPs in the intertwined lattice.
It should be noted that three symmetry-protected EPs exist in the twofold $\mathcal{PT}$-symmetric ladder lattice: two surface EPs arising from the twofold $\mathcal{PT}$-symmetry, and a bulk EP from the $\mathcal{PT}$-symmetric lattice itself.

We now propose a two-dimensional (2D) lattice in which a dissipationless one-dimensional (1D) waveguide is realizable by repeating 1D twofold $\mathcal{PT}$-symmetric ladder lattices.
For example, a straight waveguide is considered on the 2D lattice as the interface along the y-axis, as shown in the lower inset of Fig.~\ref{fig4} (a). This 2D lattice satisfies both translational symmetry along the y-axis and twofold $\mathcal{PT}$-symmetry of each 1D ladder lattice. 
Figure~\ref{fig4} (a) plots the imaginary parts of the eigenenergies as a function of $\gamma$ in the 2D bilayered square lattice with $40 \times 40$ unit cells, where it is apparent that the imaginary eigenenergies are broadened when compared to those of the 1D lattice in Fig.~\ref{fig2} (d) because of finite-size effects.
As shown in the upper inset of Fig.~\ref{fig4} (a), the distribution of the complex eigenenergies is similar to that of the 1D lattice, except for the real energy distribution due to the dispersion toward the y-axis. 
The zero-dimensional localized NHBSs, which are robust across a wide range of non-Hermiticity in the 1D lattice, extend into the 1D NHBSs in the 2D lattice. The corresponding eigenstates, which extend along the $y$ direction with no dissipation along the x-axis, are depicted in Fig.~\ref{fig4} (b), with wavelengths related to the real parts of the eigenenergies. 

Finally, we show the robustness of the dissipationless 1D NHBSs against geometrical deformations. We design a 2D $\mathcal{PT}$-symmetric layer composed of square lattices that contain a locally inverted area with an arbitrary shape, as shown in Fig.~\ref{fig4} (c). 
Notably, the geometry presents many abrupt corners where the interface bends but still conserves symmetry. We can see that the NHBSs survive without dissipation at the interfaces with sharp corners, as shown in Fig.~\ref{fig4} (d), while there exists bending loss around the sharply deformed region of a trivial waveguide. Therefore, NHBSs can be implemented, without bending loss, as the effective waveguide in a 2D lattice while conserving local twofold $\mathcal{PT}$-symmetry.

In conclusion, we proposed a system in which localized states emerge solely through the non-triviality resulting from non-Hermiticity. Our twofold $\mathcal{PT}$-symmetric ladder lattice contains symmetry-protected interface states as NHZMs and NHBSs with corresponding phase transitions at the two surface EPs and bulk EP as a function of $\gamma$.
The characteristics of these two states are as follows. The NHZMs protected by NHPH symmetry are localized at the interface between the odd surface EP and the bulk EP, while the NHBSs protected by $\mathcal{PT}$-symmetry are localized at the interface between the even surface EP and the bulk EP.
Both have constant localization lengths unaffected by changes in the non-Hermiticity parameter. As an example, we showed here that the NHBSs can form an effective waveguide without dissipation at the interface between non-Hermitian time-reversal partners.
We expect the characteristics of these two symmetry-protected interface states to open up a new field of synthetic non-Hermitian systems. 

\section*{Acknowledgments}

This work was supported by Project code (IBS-R024-D1), the National Research Foundation of Korea (NRF) grant (NRF2016-R1D1A1B04-935798) and Korea Institute for Advanced Study(KIAS) funded by the Korea government (MSIT).


\clearpage
\newpage

\pagebreak
\widetext
\begin{center}
\textbf{\large Supplemental Materials: Emergent localized states through intertwined non-Hermitian lattice}
\end{center}

\section{Hamiltonian of a ladder lattice}

\begin{figure*}
\begin{center}
\includegraphics[width=\figsizethree\textwidth]{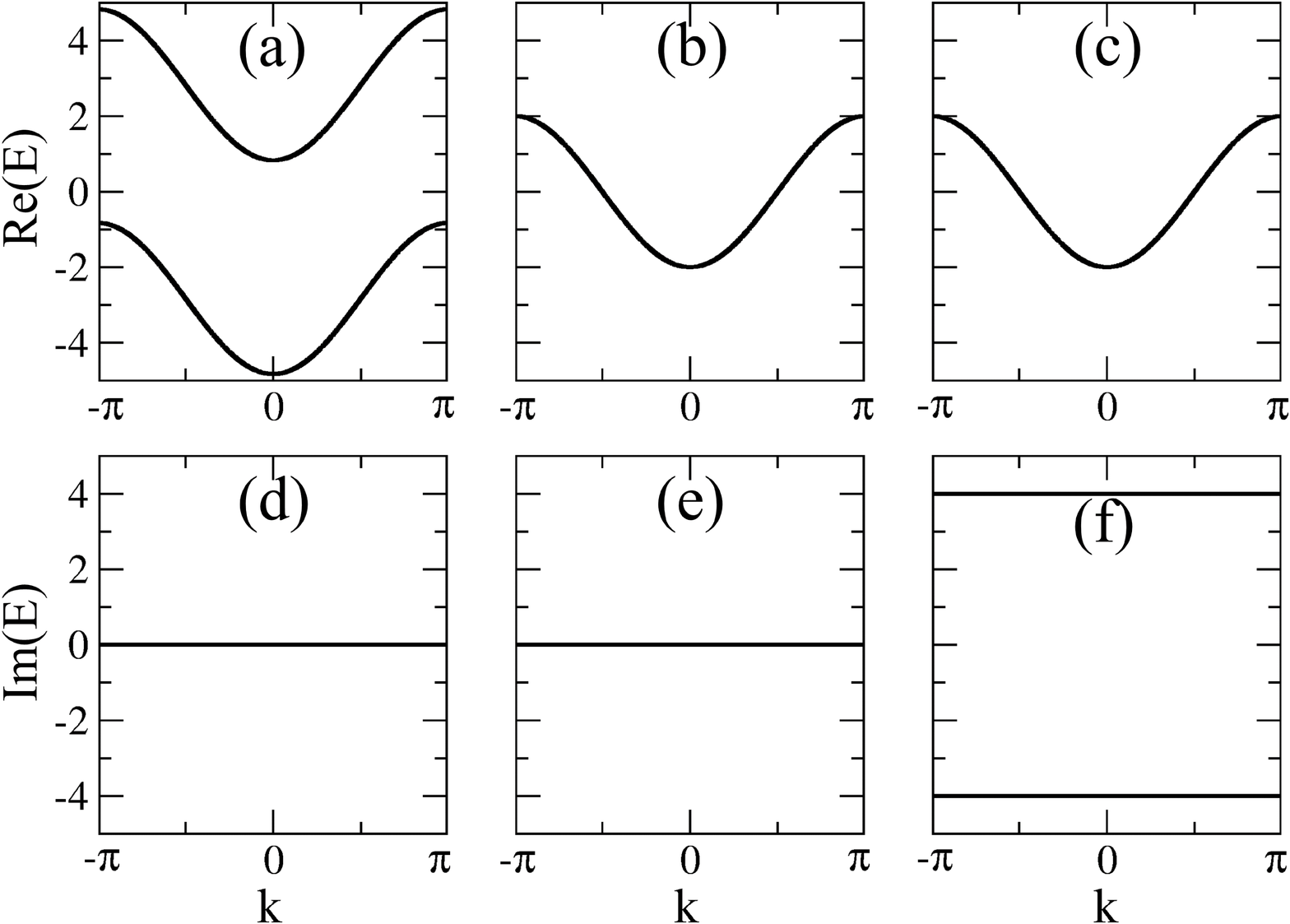}
\caption{(color online).
(a--c) Real and (d--f) imaginary parts of the complex energy bands of PT-symmetric ladder lattices with (a, d) $\gamma = 2.0$, (b, e) $\gamma = 6.0$, and (c, f) $\gamma = 10.0$.
}
\label{figs1}
\end{center}
\end{figure*}

The Hamiltonian of the ladder lattice is given by
\begin{equation}
E \Psi_j = H_0 \Psi_j + H_1 \Psi_{j+1} + H_1^{+} \Psi_{j-1},
\end{equation}
where
\begin{equation}
H_0 = 
\begin{pmatrix}
\epsilon_a  & -d \\
-d & \epsilon_b
\end{pmatrix},
H_1 = 
\begin{pmatrix}
-t  & 0 \\
0 & -t
\end{pmatrix},
\end{equation}
and $\Psi_j = (\phi_j^a, \phi_j^b)^T$. We can set $\Psi_{j+1} = \Psi_{j} e^{i k}$ and $\Psi_{j-1} = \Psi_{j} e^{-i k}$ due to the translational symmetry of the unit cells. Finally,
\begin{equation}
H = 
\begin{pmatrix}
\epsilon_a - 2 t \cos k  & -d \\
-d & \epsilon_b - 2 t \cos k
\end{pmatrix}.
\end{equation}
Solving the eigenproblem of $H$ when $\epsilon_a = -\epsilon_b = i \gamma / 2$, we obtain the band structure for the PT-symmetric ladder lattice shown in the inset of Fig.~\ref{fig2} (a). Figure~\ref{figs1} shows the real and imaginary parts of the complex energy bands with $d=3$ and $t=1$ when $\gamma = 2.0$, $6.0$, and $10.0$.

\section{Numerical method for eigenenergies and eigenstates in the twofold PT-symmetric ladder lattice}

\begin{figure*}
\begin{center}
\includegraphics[width=\figsizethree\textwidth]{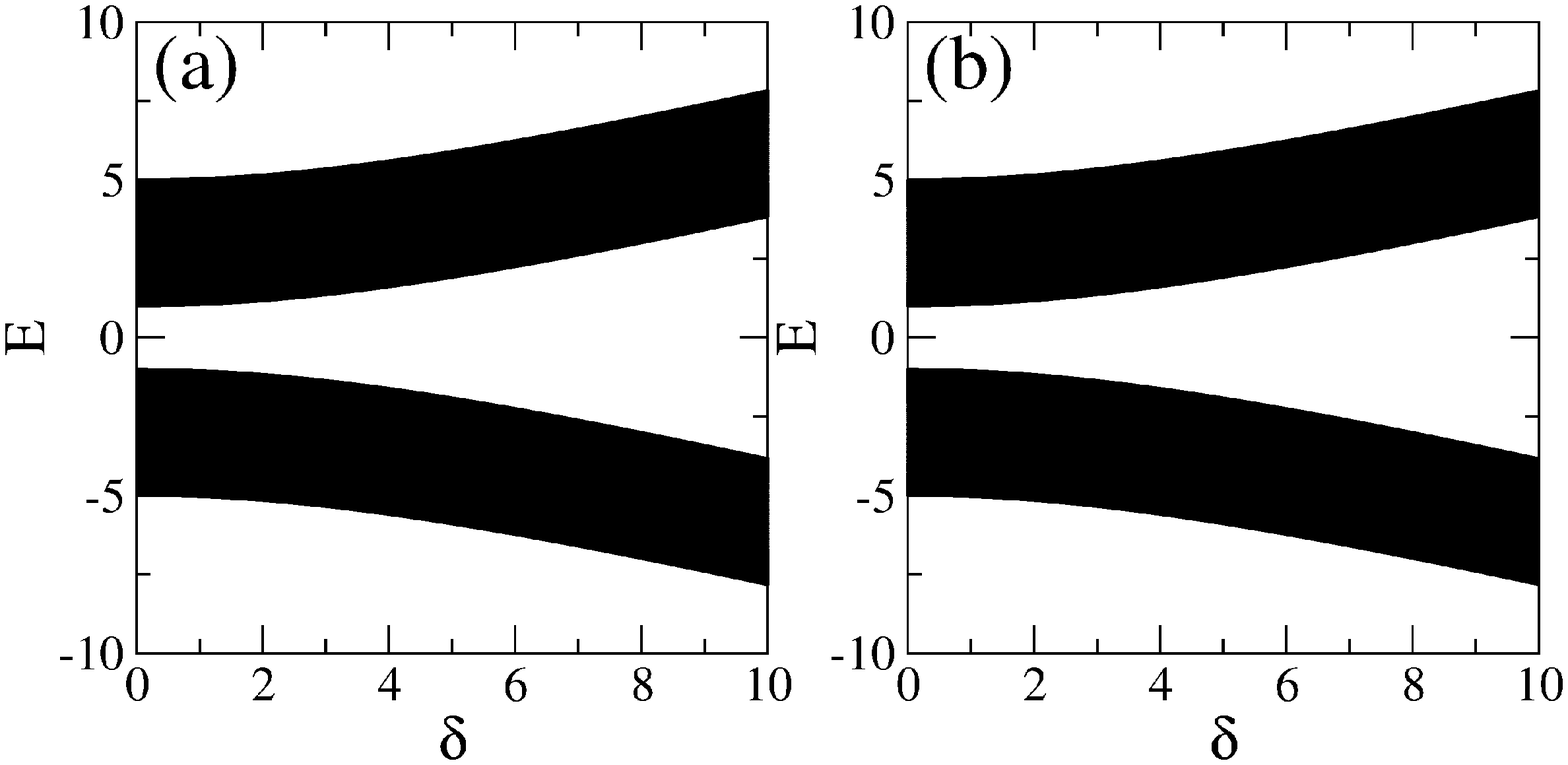}
\caption{(color online).
(a) Real eigenenergies as a function of $\delta$ when $\epsilon_a = -\epsilon_b = \delta$, $d=3$, $t=1$, and $N=100$ in a ladder lattice of the inset in Fig.~\ref{fig2} (a). (b) Real eigenenergies as a function of $\delta$ when $\epsilon_a = -\epsilon_b = \delta$, $d=3$, $t=1$, and $N=100$ in a twisted ladder lattice of the inset in Fig.~\ref{fig2} (c).
}
\label{figs2}
\end{center}
\end{figure*}

We numerically obtain the eigenenergies and eigenstates in a finite-sized twofold PT-symmetric ladder lattice with $N$ unit cells given by
\begin{equation}
\begin{pmatrix} 
\ddots &         &        &          &         &.           \\
           & H_0 & H_1 &          &         &            \\
           & H_1^{+} & H_0 & H_1  &         &            \\
           &        & H_1^{+} & H'_0 & H_1  &            \\
           &        &         & H_1^{+} & H'_0 &            \\
           &        &         &        &          & \ddots 
\end{pmatrix},
\end{equation}
where
\begin{eqnarray}
H'_0 &=& 
\begin{pmatrix} 
\epsilon_a  & -d \\
-d & \epsilon_b
\end{pmatrix} \quad\quad \text{for a ladder lattice, and} \\
H'_0 &=& 
\begin{pmatrix} 
\epsilon_b  & -d \\
-d & \epsilon_a
\end{pmatrix} \quad\quad \text{for a twisted ladder lattice.}
\end{eqnarray}
Solving this $2N \times 2N$ matrix, we can obtain $2N$ eigenenergies. For instance, in the Hermitian case of $\epsilon_a = -\epsilon_b = \delta/2$ ($\delta$ is real), the eigenenergies as functions of $\delta$ are shown in Fig.~\ref{figs2}.

\section{Analytic solution of the interface states}

\begin{figure*}
\begin{center}
\includegraphics[width=\figsizethree\textwidth]{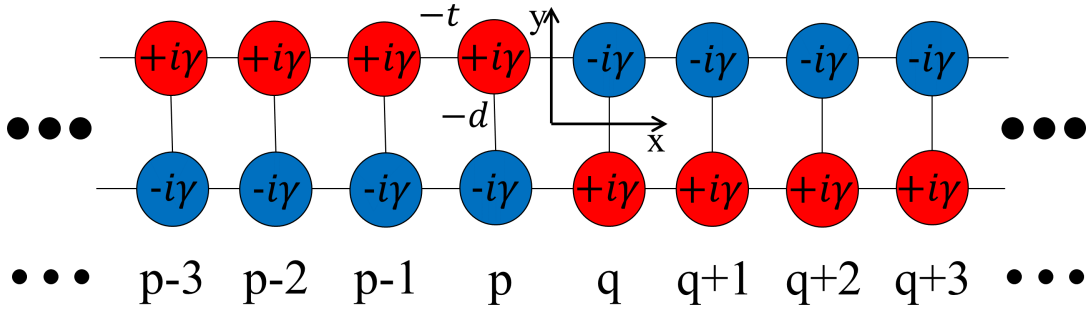}
\caption{(color online). An intertwined $\mathcal{PT}$-symmetric ladder lattice we considered.
}
\label{figs3}
\end{center}
\end{figure*}

The left and right lattices of Fig.~\ref{figs3} can be described by
\begin{eqnarray}
H \ket{\psi}_L = E_{\pm} (k) \ket{\psi}_L - t e^{i p k} (a \ket{p+1,A} + b \ket{p+1,B}) + t e^{i (p+1) k} (a \ket{p,A} + b \ket{p,B}), \\
H \ket{\psi}_R = E_{\pm} (k) \ket{\psi}_R - t e^{i q k} (b \ket{q-1,A} + a \ket{q-1,B}) + t e^{i (q-1) k} (b \ket{q,A} + a \ket{q,B}),
\end{eqnarray}
where
\begin{eqnarray}
H &=& \Sigma_n [i \gamma (\ket{n,A}\bra{n,A}-\ket{n,B}\bra{n,B}) \\
& & - d (\ket{n,A}\bra{n,B} + \ket{n,B}\bra{n,A}) -t (\ket{n+1,A}\bra{n,A}-\ket{n+1,B}\bra{n,B} + h.c.)], \\
E_{\pm}(k) &=& - 2 t \cos{k} \pm \sqrt{d^2 - (\gamma/2)^2},\\
\ket{\psi}_L &\equiv& \Sigma_{n=-\infty}^{p} e^{i n k} (a \ket{n,A} + b \ket{n,B}),\\
\ket{\psi}_R &\equiv& \Sigma_{n=q}^{\infty} e^{-i n k} (b \ket{n,A} + a \ket{n,B}),\\
a_{\pm} &=& - i \gamma/2 \mp \sqrt{d^2 - (\gamma/2)^2},\\
b_{\pm} &=& d.
\end{eqnarray}
Here, $\cos{k_{\pm}} = \frac{-E \pm \sqrt{d^2 + (\gamma/2)^2}}{2 t}$ and $\sin{k_{\pm}} = \sqrt{1-\cos^2{k_{\pm}}}$ for $L$, while $k_{\pm} \rightarrow - k_{\pm}$ for $R$. The ansatz is $\ket{\Phi} = \ket{\Phi}_{L} + \ket{\Phi}_{R}$, where $\ket{\Phi}_{L} = \alpha_{+} \ket{\psi(k_+)}_{L} + \alpha_{-} \ket{\psi(k_-)}_{L}$ and $\ket{\Phi}_{R} = \beta_{+} \ket{\psi(- k_+)}_{R} + \beta_{-} \ket{\psi(-k_-)}_{R}$. We find the condition satisfying $H \ket{\Phi} = E \ket{\Phi}$ as
\begin{equation}
2 d^2 \sin{(k_{-})} \sin{(k_{+})} + \left(\frac{\gamma}{2}\right)^2 \cos{(k_+ + k_-)} = \left(\frac{\gamma}{2}\right)^2.
\label{con1}
\end{equation}
This is derived by
\begin{eqnarray}
(-\alpha_+ a_+ e^{-i k_+} - \alpha_- a_- e^{-i k_-} + \beta_+ b_+ e^{i k_+} + \beta_- b_- e^{i k_-}) \ket{q, A} +
(\alpha_+ a_+ + \alpha_- a_- - \beta_+ b_+ - \beta_- b_-) \ket{p, A} && \\\nonumber
+(-\alpha_+ b_+ e^{-i k_+} - \alpha_- b_- e^{-i k_-} + \beta_+ a_+ e^{i k_+} + \beta_- a_- e^{i k_-}) \ket{q, B} +
(\alpha_+ b_+ + \alpha_- b_- - \beta_+ a_+  - \beta_- a_-) \ket{p, B} &=& 0.
\end{eqnarray}
Then, the square matrix below should vanish for non-trivial solutions:
\begin{equation}
\begin{pmatrix} 
-a_+ e^{-i k_+} & -a_- e^{-i k_-} & b_+ e^{i k_+} & b_- e^{i k_-} \\
a_+ & a_-  & -b_+ & -b_- \\
-b_+ e^{-i k_+}   & -b_- e^{-i k_-} & a_+ e^{i k_+}   & a_- e^{i k_-}\\
b_+  & b_- & -a_+ & -a_-
\end{pmatrix}
\begin{pmatrix} 
\alpha_+\\
\alpha_-\\
\beta_+\\
\beta_-
\end{pmatrix}
=
\begin{pmatrix} 
0\\
0\\
0\\
0
\end{pmatrix}.
\end{equation}

By putting $\cos{k_{\pm}} = \frac{-E \pm \sqrt{d^2 + (\gamma/2)^2}}{2 t}$ and $\sin{k_{\pm}} = \sqrt{1-\cos^2{k_{\pm}}}$ into Eq.~(\ref{con1}), then the energy ($E$) of the interface states can be evaluated as a function of $\gamma$ by
\begin{eqnarray}
E_{\pm}^I = \pm \sqrt{\left(1-\frac{2 t}{d}\right)\left(d^2 - 2 d t - \left(\frac{\gamma}{2}\right)^2\right)}, \label{a_inner_ev}\\
E_{\pm}^O = \pm \sqrt{\left(1+\frac{2 t}{d}\right)\left(d^2 + 2 d t - \left(\frac{\gamma}{2}\right)^2\right)}, \label{a_outer_ev}
\end{eqnarray}
where $E_{\pm}^I$ and $E_{\pm}^O$ are the eigenenergies of the eigenstates with odd and even parities about the interface, respectively. Putting Eq.~(\ref{a_inner_ev}) into $\cos{k_{\pm}} = \frac{-E \pm \sqrt{d^2 - (\gamma/2)^2}}{2 t}$, complex wave numbers $k_{\pm}^I$ of the interface states are given by
\begin{equation}
\cos{k_{\pm}^I} = \frac{\mp \sqrt{\left(1-\frac{2 t}{d}\right)\left(d^2 - 2 d t - \left(\frac{\gamma}{2}\right)^2\right)} \pm \sqrt{d^2 - (\frac{\gamma}{2})^2}}{2 t}.
\end{equation}
This $k_{\pm}^I$ can be classified into three phases by the two transition points $\gamma_I = 2 \sqrt{d^2 - 2 d t}$ and $\gamma_b = 2 d$. Setting $k_{\pm}^I = {\tilde k}_{\pm}^I + i \kappa_{\pm}^I$, the imaginary parts $\kappa_{\pm}^I$ of complex wave number $k_{\pm}^I$, which are the reciprocals of the localization lengths of the interface states, are 
\begin{eqnarray}
\kappa_{\pm}^I =& \cosh^{-1}{\frac{ -\sqrt{\left(1-\frac{2 t}{d}\right)\left(d^2 - 2 d t - \left(\frac{\gamma}{2}\right)^2\right)} + \sqrt{d^2 - (\frac{\gamma}{2})^2}}{2 t}}   \quad\quad &\text{for $\gamma < \gamma_I$,} \\
\kappa_{\pm}^I =& \cosh^{-1}\left({\sqrt{\frac{d}{2 t}}}\right)  \quad\quad &\text{for $\gamma_I < \gamma < \gamma_b$,} \\
\kappa_{\pm}^I =& \sinh^{-1}{\frac{ - \sqrt{\left(1-\frac{2 t}{d}\right)\left(d^2 - 2 d t - \left(\frac{\gamma}{2}\right)^2\right)} + \sqrt{d^2 - (\frac{\gamma}{2})^2}}{2 t i}}   \quad\quad &\text{for $\gamma_b < \gamma$,}
\end{eqnarray}
while the real parts ${\tilde k}_{\pm}^I$ are
\begin{eqnarray}
{\tilde k}_{\pm}^I =& 0 \quad \text{or} \quad \pm \pi \quad\quad &\text{for $\gamma < \gamma_I$,} \\
{\tilde k}_{\pm}^I =& \cos^{-1}\left(\mp \sqrt{\frac{d^2 - \left(\frac{\gamma}{2}\right)^2}{2 d t}}\right)  \quad\quad &\text{for $\gamma_I < \gamma < \gamma_b$,} \\
{\tilde k}_{\pm}^I =& \mp \frac{\pi}{2} \quad\quad &\text{for $\gamma_b < \gamma$.}
\end{eqnarray}
Putting Eq.~(\ref{a_outer_ev}) into $\cos{k_{\pm}} = \frac{-E \pm \sqrt{d^2 - (\gamma/2)^2}}{2 t}$, complex wave numbers $k_{\pm}^O$ of the interface states are given by
\begin{equation}
\cos{k_{\pm}^O} = \frac{\mp \sqrt{\left(1+\frac{2 t}{d}\right)\left(d^2 + 2 d t - \left(\frac{\gamma}{2}\right)^2\right)} \pm \sqrt{d^2 - (\frac{\gamma}{2})^2}}{2 t}.
\end{equation}
Likewise, $k_{\pm}^O$ can be classified into three phases by the two transition points $\gamma_b = 2 d$ and $\gamma_O = 2 \sqrt{d^2 + 2 d t}$. Setting $k_{\pm}^O = {\tilde k}_{\pm}^O + i \kappa_{\pm}^O$, the imaginary parts $\kappa_{\pm}^O$ of complex wave number $k_{\pm}^O$, which are the reciprocals of the localization lengths of the interface states, are 
\begin{eqnarray}
\kappa_{\pm}^O =& \cosh^{-1}{\frac{ \sqrt{\left(1+\frac{2 t}{d}\right)\left(d^2 + 2 d t - \left(\frac{\gamma}{2}\right)^2\right)} - \sqrt{d^2 - (\frac{\gamma}{2})^2}}{2 t}}   \quad\quad &\text{for $\gamma < \gamma_b$,} \\
\kappa_{\pm}^O =& \cosh^{-1}\left({\sqrt{\frac{d}{2 t}+1}}\right)  \quad\quad &\text{for $\gamma_b < \gamma < \gamma_O$,} \\
\kappa_{\pm}^O =& \sinh^{-1}{\frac{-\sqrt{\left(1+\frac{2 t}{d}\right)\left(d^2 + 2 d t - \left(\frac{\gamma}{2}\right)^2\right)} + \sqrt{d^2 - (\frac{\gamma}{2})^2}}{2 t i}}   \quad\quad &\text{for $\gamma_O < \gamma$,}
\end{eqnarray}
while the real parts ${\tilde k}_{\pm}^O$ are
\begin{eqnarray}
{\tilde k}_{\pm}^O =& 0 \quad \text{or} \quad \pm \pi \quad\quad &\text{for $\gamma < \gamma_b$,} \\
{\tilde k}_{\pm}^O =& \cos^{-1}\left(\pm \sqrt{\frac{d^2 + 2 d t- \left(\frac{\gamma}{2}\right)^2}{2 d t}}\right)  \quad\quad &\text{for $\gamma_b < \gamma < \gamma_O$,} \\
{\tilde k}_{\pm}^O =& \pm \frac{\pi}{2} \quad\quad &\text{for $\gamma_O < \gamma$.}
\end{eqnarray}

\section{Hamiltonian of a 2D ladder lattice}

The Hamiltonian of the 2D ladder lattice is given by
\begin{equation}
E \Psi_{j,k} = H_0 \Psi_{j,k} + H_1 \Psi_{j+1,k} + H_1^{+} \Psi_{j-1,k} + H_2 \Psi_{j,k+1} + H_2^{+} \Psi_{j,k-1},
\end{equation}
where
\begin{equation}
H_0 = 
\begin{pmatrix}
\epsilon_a  & -d \\
-d & \epsilon_b
\end{pmatrix},
H_1 = 
\begin{pmatrix}
-t_x  & 0 \\
0 & -t_x
\end{pmatrix},
H_2 = 
\begin{pmatrix}
-t_y  & 0 \\
0 & -t_y
\end{pmatrix},
\end{equation}
and $\Psi_{j,k} = (\phi_{j,k}^a, \phi_{j,k}^b)^T$. We can set $\Psi_{j+1,k} = \Psi_{j,k} e^{i k_x}$, $\Psi_{j-1,k} = \Psi_{j,k} e^{-i k_x}$, $\Psi_{j+1,k} = \Psi_{j,k+1} e^{i k_y}$, and $\Psi_{j,k-1} = \Psi_{j,k} e^{-i k_y}$ due to the translational symmetry of the unit cells. Finally,
\begin{equation}
H = 
\begin{pmatrix}
\epsilon_a - 2 t_x \cos k_x - 2 t_y \cos k_y  & -d \\
-d & \epsilon_b - 2 t_x \cos k_x - 2 t_y \cos k_y
\end{pmatrix}.
\end{equation}


\begin{thebibliography}{150}

\bibitem{Ben98} C. M. Bender and S. Boettcher, Real Spectra in Non-Hermitian Hamiltonians Having PT Symmetry, Phys. Rev. Lett. {\bf 80}, 5243 (1998).

\bibitem{Hat96} N. Hatano, and D. R. Nelson, Localization Transitions in Non-Hermitian Quantum Mechanics, Phys. Rev. Lett. {\bf 77}, 570-573 (1996).

\bibitem{Isr19} I. Klich, Closed hierarchies and non-equilibrium steady
states of driven systems, Ann. of Phys. {\bf 404}, 66-80 (2019).

\bibitem{Elg07} R. El-Ganainy, K. G. Makris, D. N. Christodoulides, and Z. H. Musslimani, Theory of coupled optical PT-symmetric structures, Opt. Lett. {\bf 32}, 2632 (2007).
\bibitem{Rue10} C. E. R{\"u}ter, K. G. Makris, R. El-Ganainy, D. N. Christodoulides, M. Segev, and D. Kip, Observation of parity-time symmetry in optics, Nat. Phys. {\bf 6}, 192-195 (2010).
\bibitem{Mus08} Z. H. Musslimani, K. G. Makris, R. El-Ganainy, and D. N. Christodoulides, Optical Solitons in PT Periodic Potentials, Phys. Rev. Lett. {\bf 100}, 030402 (2008).
\bibitem{Mak08} K. G. Makris, R. El-Ganainy, D. N. Christodoulides, and Z. H. Musslimani, Beam Dynamics in PT Symmetric Optical Lattices, Phys. Rev. Lett. {\bf 100}, 103904 (2008).
\bibitem{Kla08} S. Klaiman, U. G{\"u}nther, and N. Moiseyev, Visualization of Branch Points in PT-Symmetric Waveguides, Phys. Rev. Lett. {\bf 101}, 080402 (2008).
\bibitem{Guo09} A. Guo, G. J. Salamo, D. Duchesne, R. Morandotti, M. Volatier-Ravat, V. Aimez, G. A. Siviloglou, and D. N. Christodoulides, Observation of P T -Symmetry Breaking in Complex Optical Potentials, Phys. Rev. Lett. {\bf 103}, 093902 (2009).


\bibitem{Reg12} A. Regensburger, C. Bersch, M. A. Miri, G. Onishchukov, D. N. Christodoulides, and U. Peschel, Parity-time synthetic photonic lattices, Nature (London) {\bf 488}, 167-171 (2012).

\bibitem{Sch11} J. Schindler, A. Li, M. C. Zheng, F. M. Ellis, and T. Kottos, Experimental study of active LRC circuits with PT symmetries, Phys. Rev. A {\bf 84}, 040101 (2011).
\bibitem{Jog10} Y. N. Joglekar, D. Scott, M. Babbey, and A. Saxena, Robust and fragile PT-symmetric phases in a tight-binding chain, Phys. Rev. A {\bf 82}, 030103(R) (2010).
\bibitem{Laz13} N. Lazarides and G. P. Tsironis, Phys. Gain-Driven Discrete Breathers in PT-Symmetric Nonlinear Metamaterials, Rev. Lett. {\bf 110}, 053901 (2013).

\bibitem{Kat96} T. Kato, {\it Perturbation Theory of Linear Operators} (Springer, Berlin, 1996).
\bibitem{Hei12} W. D. Heiss, The physics of exceptional points. J. Phys. A {\bf 45}, 444016 (2012).

\bibitem{Dem01} C. Dembowski, H.-D. Gr{\"a}f, H. L. Harney, A. Heine, W. D. Heiss, H. Rehfeld, and A. Richter, Experimental Observation of the Topological Structure of Exceptional Points, Phys. Rev. Lett. {\bf 86}, 787 (2001).
\bibitem{Dem03} C. Dembowski, B. Dietz, H.-D. Gr{\"a}f, H. L. Harney, A. Heine, W. D. Heiss, and A. Richter, Observation of a Chiral State in a Microwave Cavity, Phys. Rev. Lett. {\bf 90}, 034101 (2003).
\bibitem{Lee09} S.-B. Lee, J. Yang, S. Moon, S.-Y. Lee, J.-B. Shim, S. W. Kim, J.-H. Lee, and K. An, Observation of an Exceptional Point in a Chaotic Optical Microcavity, Phys. Rev. Lett. {\bf 103}, 134101 (2009).

\bibitem{Gao15} T. Gao, E. Estrecho, K. Y. Bliokh, T. C. H. Liew, M. D. Fraser, S. Brodbeck, M. Kamp, C. Schneider, S. H{\"o}fling, Y. Yamamoto, F. Nori, Y. S. Kivshar, A. G. Truscott, R. G. Dall, and E. A. Ostrovskaya, Observation of non-Hermitian degeneracies in a chaotic exciton-polariton billiard, Nature {\bf 526}, 554-558 (2015).

\bibitem{Li15} L. Ge, Parity-time symmetry in a flat-band system, Phys. Rev. A {\bf 92}, 052103 (2015).
\bibitem{Li18} L. Ge, Non-Hermitian lattices with a flat band and polynomial power increase, Photon. Res. {\bf 6}, A10 (2018).

\bibitem{LGe17} L. Ge, Symmetry-protected zero-mode laser with a tunable spatial profile, Phys. Rev. A {\bf 95}, 023812 (2017).
\bibitem{Qi18} B. Qi, L. Zhang, and L. Ge, Defect States Emerging from a Non-Hermitian Flatband of Photonic Zero Modes, Phys. Rev. Lett. {\bf 120}, 093901 (2018).

\bibitem{Mal15} S. Malzard, C. Poli, and H. Schomerus, Topologically Protected Defect States in Open Photonic Systems with Non-Hermitian Charge-Conjugation and Parity-Time Symmetry, Phys. Rev. Lett. {\bf 115}, 200402 (2015).
\bibitem{Mal18} S. Malzard and H. Schomerus, Bulk and edge-state arcs in non-Hermitian coupled-resonator arrays, Phys. Rev. A {\bf 98}, 033807 (2018).

\bibitem{Sch13} H. Schomerus, Topologically protected midgap states in complex photonic lattices, Opt. Lett. {\bf 38}, 1912-1914 (2013).
\bibitem{Pol15} C. Poli, M. Bellec, U. Kuhl, F. Mortessagne, and H. Schomerus, Selective enhancement of topologically induced interface states in a dielectric resonator chain, Nat. Commun. {\bf 6}, 6710 (2015).
\bibitem{Zeu15} J. M. Zeuner, M. C. Rechtsman, Y. Plotnik, Y. Lumer, S. Nolte, M. S. Rudner, M. Segev, and A. Szameit, Observation of a Topological Transition in the Bulk of a Non-Hermitian System, Phys. Rev. Lett. {\bf 115}, 040402 (2015).
\bibitem{Zha18} H. Zhao, P. Miao, M. H. Teimourpour, S. Malzard, R. El-Ganainy, H. Schomerus, and L. Feng, Topological hybrid silicon microlasers, Nat. Commun. {\bf 9}, 981 (2018).
\bibitem{Par18} M. Parto, S. Wittek, H. Hodaei, G. Harari, M. A. Bandres, J. Ren, M. C. Rechtsman, M. Segev, D. N. Christodoulides, and M. Khajavikhan, Edge-Mode Lasing in 1D Topological Active Arrays, Phys. Rev. Lett. {\bf 120}, 113901 (2018). 

\bibitem{Su79} W. P. Su, J. R. Schrieffer, and A. J. Heeger, Solitons in polyacetylene, Phys. Rev. Lett. {\bf 42}, 1698 (1979).

\bibitem{Wei17} S. Weimann, M. Kremer, Y. Plotnik, Y. Lumer, S. Nolte, K. G. Makris, M. Segev, M. C. Rechtsman, and A. Szameit, Topologically protected bound states in photonic parity-time-symmetric crystals, Nat. Mater. {\bf16}, 433-438 (2017).
\bibitem{Ni18} X. Ni, D. Smirnova, A. Poddubny, D. Leykam, Y. Chong, A. B. Khanikaev, Exceptional points in topological edge spectrum of PT symmetric domain walls, Phys. Rev. B {\bf 98}, 165129 (2018).

\bibitem{Lee16} T. E. Lee, Anomalous Edge State in a Non-Hermitian Lattice, Phys. Rev. Lett. {\bf 116}, 133903 (2016).
\bibitem{Mar18} V. M. Martinez Alvarez, J. E. Barrios Vargas, and L. E. F. Foa Torres, Non-Hermitian robust edge states in one dimension: Anomalous localization and eigenspace condensation at exceptional points, Phys. Rev. B {\bf 97}, 121401(R) (2018).

\bibitem{Ley17} D. Leykam, K. Y. Bliokh, C. Huang, Y. D. Chong, and F. Nori, Edge Modes, Degeneracies, and Topological Numbers in Non-Hermitian Systems, Phys. Rev. Lett. {\bf 118}, 040401 (2017).

\bibitem{Zha15} H. Zhao, S. Longhi, and L. Feng, Robust light state by quantum phase transition in non-hermitian optical materials, Sci. Rep. {\bf 5}, 17022 (2015).
\bibitem{Pan18} M. Pan, H, Zhao, P. Miao, S. Longhi, and L. Feng, Photonic zero mode in a non-Hermitian photonic lattice, Nat. Commun. {\bf 9}, 1308 (2018).

\bibitem{LGe14} L. Ge and A. D. Stone, Parity-Time Symmetry Breaking beyond One Dimension: The Role of Degeneracy, Phys. Rev. X {\bf 4}, 031011, (2014).
\bibitem{Liu16} Z. Liu, Q. Zhang, X. Liu, Y. Yao, and J.-J. Xiao, Absence of Exceptional Points in Square Waveguide Arrays with Apparently Balanced Gain and Loss, Sci. Rep. {\bf 6}, 22711 (2016).
\bibitem{Din16} K. Ding, G. Ma, M. Xiao, Z. Q. Zhang, and C. T. Chan, Emergence, Coalescence, and Topological Properties of Multiple Exceptional Points and Their Experimental Realization, Phys. Rev. X, {\bf 6}, 021007 (2016).

\bibitem{She18} H. Shen, B. Zhen, and L. Fu, Topological Band Theory for Non-Hermtian Hamiltonian, Phys. Rev. Lett. {\bf 120}, 146402 (2018).


\end{thebibliography}
\end{document}